\documentclass[fleqn,twoside]{article}
\usepackage{espcrc2}

\usepackage{graphicx}
\usepackage[figuresright]{rotating}

\newcommand{\AmS}{{\protect\the\textfont2
  A\kern-.1667em\lower.5ex\hbox{M}\kern-.125emS}}

\RequirePackage{xspace}
\usepackage{relsize}
\def\babar{\mbox{\slshape B\kern-0.1em{\smaller A}\kern-0.1em B\kern-0.1em{\smaller A\kern-0.2em R}}\xspace}
\def\superb  {Super$B$\xspace}
\def\belle{Belle\xspace}
\def\belletwo{Belle II\xspace}
\def\lhcb{LHCb\xspace}
\def\lhc{LHC\xspace}

\def\sinsqthetaw {\ensuremath{\sin^2\theta_{W}}\xspace}

\def\invab   {\ensuremath{\mbox{\,ab}^{-1}}\xspace}
\def\invfb   {\ensuremath{\mbox{\,fb}^{-1}}\xspace}

\def\FourS {\ensuremath{\Upsilon{(4S)}}\xspace}
\def\FiveS {\ensuremath{\Upsilon{(5S)}}\xspace}
\def\NS {\ensuremath{\Upsilon{(nS)}}\xspace}

\def\CP {\ensuremath{CP}\xspace}

\newcommand\vub {\ensuremath{V_{ub}}}

\newcommand\vcb {\ensuremath{V_{cb}}}
\newcommand\vtd {\ensuremath{V_{td}}}
\newcommand\vts {\ensuremath{V_{ts}}}

\title{Super Flavour Factory Round Table -- The Super$B$ Physics Programme}

\author{A. J. Bevan\address{Department of Physics, Queen Mary, University of London, Mile End Road, London, E1 4NS, UK.}
}%
       
\begin{document}

\begin{abstract}
\superb is a proposed high luminosity Super Flavour Factory capable of accumulating
75\invab of data at the \FourS as well as at other center of mass energies.
These proceedings summarise highlights of the \superb physics programme, and 
in particular there is emphasis on the unique aspects of \superb over other 
planned or existing experiments.
\vspace{1pc}
\end{abstract}

\maketitle

\section{INTRODUCTION}

\superb is a proposed high luminosity $e^+e^-$ collider with a design luminosity of 
$10^{36} cm^{-2} s^{-1}$.  This experiment will accumulate 75\invab of data at the \FourS
with five nominal years of data taking, which is approximately 65 times the combined
\babar and \belle data sample at this energy.  In addition to accumulating data at
the \FourS, \superb will run at energies from charm threshold, $\psi(3770)$, to
the $\Upsilon(6S)$.  The $e^+e^-$ collider will have an 80\% polarised electron beam,
that also has significant impact on the physics potential of this experiment.  The
broad physics programme of \superb is described in Ref.~\cite{physicswp}.
The \superb accelerator complex is described in~\cite{panta_2006,ref:white_acc}, and the 
detector concept is reviewed in~\cite{detectorwp}.

While the baseline luminosity is $10^{36} cm^{-2} s^{-1}$, if the accelerator is able to
achieve this performance with the nominal lattice parameters, there is a significant 
potential to upgrade the luminosity by a factor of four over the lifetime of the experiment.

\superb is a natural successor to the \babar, \belle, and BES-III experiments, as it will accumulate about
two orders of magnitude data more than these experiments will have delivered during
their lifetimes. There is competitor experiment to \superb, called \belletwo, which is
in the early stages of being constructed and aims to accumulate about 50\invab  
during its lifetime.  The physics potential of \belletwo can be found
in Refs.~\cite{belletworpt,belletwo}.  The main advantages of \superb over other experiments
are discussed in Section~\ref{bevan:physicsoverview}.
The remainder of these proceedings discuss
highlights of the $B$ physics programme (Section~\ref{bevan:bphysics}), 
the potential of Charm Physics in general and at the $\psi(3770)$ (Section~\ref{bevan:charm}),
precision electroweak measurements, in particular the potential to measure \sinsqthetaw
(Section~\ref{bevan:precisionew}), benefits of beam polarisation to the $\tau$
physics programme (Section~\ref{bevan:tau}), and the potential for direct searches and 
SM spectroscopy studies (Section~\ref{bevan:spectroscopy}).  A difficult aspect of the 
physics programme of \superb is the phenomenological  archaeology that will be required to 
try and elucidate new physics should it be manifest in the data.  Our current understanding of
how this may work is summarised in Section~\ref{bevan:archiologoy}. It should be noted that these
sections concentrate mostly on the unique features of the \superb experiment as examples of the
wider programme.  
The final section of these proceedings provides
a brief summary of the highlights of this work.

\section{Overview of the Physics Programme}
\label{bevan:physicsoverview}

The physics programme of \superb can be summarised as the search for direct and indirect
signs of physics beyond the Standard Model (SM), generically referred to as {\em new physics},
while simultaneously performing precision tests of the SM~\cite{physicswp}.  The searches for new physics are
sensitive to particles with masses far in excess of the reach of \lhc experiments, Lepton Flavour Violating (LFV)
processes in the $\tau$ sector, on mass shell light dark matter and light Higgs candidates, and
manifestations of so-called Dark Forces.  
More detail on the complete programme can be found in Refs.~\cite{physicswp}
and~\cite{chang} which contain details of the physics programme common to both \belletwo and \superb.  
A theoretical overview of the physics case for Super Flavour Factories can be found
in Ref.~\cite{marco}.

The following aspects of this programme are unique:
\begin{itemize}
 \item A larger baseline data set than any proposed experiment at these energies.  With the additional data that
   \superb aims to integrate within five years of nominal running, one should be able to observe
   several rare decays that are sensitive to new physics, if those decays occur at the expected
   SM rate.  These rare decays may play an important role in constraining details of the new physics Lagrangian.
 \item The ability to run at the $\psi(3770)$ which corresponds to charm threshold.  This opens 
   up the possibility to study time-dependent \CP asymmetries in $D^0\overline{D}^0$ decays in
   analogy to the \CP violation studies that have been done at the existing $B$ factories.  In
   addition to this unique potential, by accumulating a large sample of data at the $\psi(3770)$,
   one will be able to make precision measurements of a number of decay constants and other parameters.
   These will improve theoretical understanding of experimental programmes at LHCb and the 
   Super Flavour Factories, in particular the measurements of $\gamma$, and of charm mixing.  
   Many of these measurements
   will be useful in validating Lattice QCD and theoretical frameworks, and in parallel many rare decay studies
   will be able to provide constraints on new physics.
 \item By having a polarised electron beam, it is possible to make precision tests of the electroweak
   sector that complement SLC and LEP measurements of $\sin^2\theta_W$.  The benefit of having such a measurement is that it is
   essentially free from theoretical uncertainties. The 
   polarised electron beam will enable one to reconstruct the $\tau$ polarisation, and use this 
   information as a discriminating variable when searching for lepton flavour violating processes.  One will also
   be able to measure the $\tau$ EDM and $g-2$ using these data.
\end{itemize}

Once the design goals of \superb have been achieved, there is scope to increase luminosity by up to 
a factor of four.  If that is realisable then, just as with the $B$ factories, the physics programme
of \superb will expand significantly.  Those measurements of rare processes or small effects that would
have been marginal in terms of sensitivity with 75\invab would provide very significant constraints to further
our understanding of nature at high energy if one accumulates several hundred \invab.  The precision of 
measurements that are central to the physics programme of this experiment would also be surpassed
in almost all circumstances. As with any 
frontier, by pushing the intensity to a new level, one would be exposed to additional opportunities
to constrain nature.  

\superb will be able to cover a wider range of measurements than the \belletwo experiment, with
more data. Where there is overlap between the programmes of these two experiments, one can expect a 
repeat of the excellent synergies that existed between the LEP experiments, \babar and \belle, and 
will no doubt be present at the LHC in coming years.  The dedicated flavour experiment at the \lhc, 
called \lhcb, will mostly probe complementary flavour observables to \superb.  In the few cases 
where there is overlap between these experimental programmes, again that will provide a useful cross 
check of performance, and as noted above, measurements from \superb will play an important role
in controlling theoretical uncertainties associated with the interpretation of some of the results from
\lhcb.

Measurements from \superb will have ramifications for both fundamental particle physics and cosmology.  Many of these
results will help us understand the flavour sector of the SM and new physics scenarios, which 
in turn may have relevance for the matter-antimatter asymmetry problem, and the origin of the Universe. Studies
of rare decays may help elucidate the type of new physics and energy scale that this occurs at. 
Precision measurements of $\sin^2\theta_W$, are related to the electroweak symmetry
breaking (EWSB) process, central to the SM, and searches for light Higgs particles may elucidate EWSB beyond the SM. 
In addition to these, one can elucidate Dark Matter and Dark Forces 
postulates, as well as probing the effects
of quantum gravity through precision tests of CPT in $B$, $D$, and $\tau$ decays.  Tests of other fundamental
symmetries, such as lepton universality may also yield a surprise.

Table~\ref{tbl:superbmeas} gives a summary of expectations for some of the main measurements to be made at \superb.
The following sections discuss some of the unique features of the 
physics programme of this experiment in more detail.

\begin{table}[!ht]
\caption{A summary of expected precision obtained for a number of benchmark channels at \superb.  See Ref.~\cite{physicswp}
for a more comprehensive overview. $B_s$ measurements assume 1 (30)\invab of data at the \FiveS, and 
the charm prospects assume that in addition to \FourS data, one has accumulated a sample of 500\invfb of
data at $\psi(3770)$.}\label{tbl:superbmeas}
\begin{tabular}{lc}
\hline
Measurement & Precision \\ \hline
B Decays & \\
$B\to K\nu \overline{\nu}$   & observe \\
$B\to K^*\nu \overline{\nu}$ & observe \\
$\Delta S (\eta^\prime K^0)$ & $\pm 0.02$ \\
$\beta (c\overline{c}s)$      & $0.1^\circ$\\
$\alpha$                     & $1-2^\circ$\\
$\gamma$                     & $1-2^\circ$\\
$A^s_{SL}$                   & 0.006 (0.004)\\
$B_s \to \gamma\gamma$       & 38\% (7\%) \\

\hline

Charm & \\
$x_D$ & $\pm 2.0\times 10^{-4}$\\
$y_D$ & $\pm 1.2\times 10^{-4}$\\
\hline

Precision Electroweak & \\
$\sin^2 \theta_W$ ($\sqrt{s} = m_{\FourS}$)    & 1\% \\

\hline
$\tau$ Physics        & \\
$\tau \to \mu\gamma$     & ${\cal{B}} < 2.4\times 10^{-9}$ \\
$\tau \to e\gamma$       & ${\cal{B}} < 2.4\times 10^{-9}$ \\
$\tau \to \ell\ell\ell$  & ${\cal{B}} < 2 - 8\times 10^{-10}$ \\ \hline
\end{tabular}
\end{table}

\section{$B$ Physics}
\label{bevan:bphysics}

\begin{figure}[!ht]
\resizebox{8cm}{!}{
\includegraphics[width=0.44\textwidth]{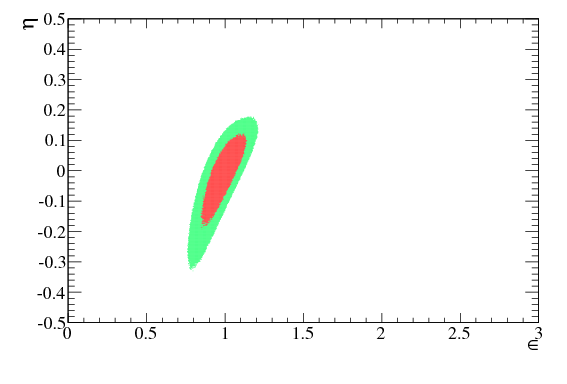}
}
        \caption{Expected constraints the new physics parameters $\epsilon$ and $\eta$ using $B\to K^{(*)}\nu\overline{\nu}$ 
                 decays at \superb (not including the measurement of $f_{L}$ from the $K^*$ mode).  The SM solution
        corresponds to the point $(\epsilon, \eta) = (1, 0)$.} \label{fig:bevan:snunubar}
\end{figure}

A number of rare $B$ decays are sensitive to new physics through loops or Flavour Changing
Neutral Current (FCNC) processes.  In particular decays with suppressed SM amplitudes, that 
could interfere with significant non-SM amplitudes could be sensitive 
probes of new physics.  There are a number of such channels, with interesting observables
ranging from branching fractions and forward-backward asymmetries, to 
time-dependent CP asymmetry parameters discussed in Ref.~\cite{physicswp}.  One particular example that is a golden 
channel for \superb is $B\to K^{(*)}\nu\overline{\nu}$.  In order to observe 
these decays occurring at a SM rate, one needs to accumulate of the order of 75\invab of 
data.  With such an observation it would be possible to measure the branching
fractions of both $B\to K\nu\overline{\nu}$ and $B\to K^{*}\nu\overline{\nu}$,
as well as the fraction of longitudinally polarised events $f_L$ in the latter mode.
An experiment with a smaller data sample would not be able to measure all of these observables.
The corresponding constraint obtained on the new physics parameters $\epsilon$ and $\eta$
(see Ref.~\cite{Altmannshofer:2009ma}) are shown in Figure~\ref{fig:bevan:snunubar}.

While many of the measurements at \superb will focus on detailed studies of CP violation,
one should not neglect the possibility that CPT is violated.  This symmetry, while conserved
in the SM as a result of the intrinsic Lorentz structure of the model, can be violated
in scenarios beyond the SM, for example~\cite{Kostelecky:1997mh}.  \superb will be able to produce more stringent constraints on
CPT in $B^0\overline{B}^0$ oscillations than previous experiments, reaching a precision 
of $0.3-0.6$ per mille on the real and imaginary parts of the CPT violating mixing parameter $z$.

In addition to the aforementioned searches for new physics, one should also recall
that it will be possible to perform precision tests of the CKM mechanism using both
direct and indirect constraints on the unitarity triangle.  The observables accessible
to a Super Flavour Factory include several different measurements of all of the angles 
($\alpha$, $\beta$, $\gamma$)\footnote{The alternate notation of ($\phi_1$, $\phi_2$, $\phi_3$) 
= ($\beta$, $\alpha$, $\gamma$) may also found in the literature.},
as well as measurements of \vub, \vcb, \vts, and \vtd. No other type of flavour
experiment is able to perform such a set of measurements.  For this reason, both 
\superb and \belletwo will provide powerful set of precision constraints on the description
of CP violation and quark mixing in the SM. 
Ref.~\cite{physicswp} contains more
details on this part of the \superb physics programme.

\section{\boldmath Charm Physics}
\label{bevan:charm}

The discovery of mixing in neutral $D$ mesons has a profound implication on the phenomenology of
charm decays.  As with neutral kaons and $B$ mesons, the establishment of mixing, which is 
interesting in its own right, also brings the potential for many new CP violating observables
to be studied at future experiments.  Precision measurements of $D$ mixing will be possible
at \superb, in addition to searching for CP violation in $D$ mesons. Such measurements 
would be the only probes of CP asymmetries involving transitions of an up-type quark.
CP violation in charm decays within the SM is expected to be a small effect, as the CKM matrix
elements involved in $c$ quark transitions are mostly real, where imaginary components related to the 
CKM phase only become apparent at order $\lambda^4$~\cite{bevan:wolfenstein}.
If the CKM scenario is able to accurately predict CP violation phenomena in the charm
sector, then this will strengthen the case that this is indeed the dominant description
of quark mixing in the SM.  Any deviation from SM expectations would yield a clear 
signature for NP.

Using only data from the \FourS, one would expect to be able to measure mixing parameters
in the charm sector $x_D$ and $y_D$ to a precision of $4.2\times 10^{-4}$, and $1.7\times 10^{-4}$,
assuming realistic input of information on strong phase differences from BES III or
using data collected by \superb at the $\psi(3770)$.  More details
of this estimate can be found in Ref.~\cite{physicswp}.

\subsection{\boldmath Charm Physics at the $\psi(3770)$}
\label{bevan:charmthreshold}

A sample of 500\invfb of data (50 times the data sample expected at the BES-III experiment) 
could be accumulated at the $\psi(3370)$ over a period of 
a few months using \superb.  The applications of this data are far-reaching, and go 
beyond the current CLEOc and BES-III programmes at the $\psi(3770)$.
One advantage of studying charm decays at low energy is the relative lack of background,
when compared to data accumulated at a higher energy resonance such as the \FourS.  This
is clearly manifest through the competitiveness of CLEOc in a number of measurements of 
charm decays, when compared to the results from \babar and \belle on a number of 
branching fraction and decay constant measurements.  The combination of kinematic 
constraint, tagged $D$ mesons, and a clean experimental environment make the $\psi(3770)$
a versatile laboratory to test many aspects of the SM and search for new physics.

Ever since the $B$ factories established mixing in the $D^0\overline{D}^0$ system, 
the possibility of utilising quantum correlations at charm threshold (the $\psi(3770)$ resonance)
for time-dependent measurements has been a possibility.  The physics sensitivity to 
CP asymmetries at charm threshold is under investigation.

Measurements of strong phases in Dalitz decays at \superb will play a significant role
in reducing theoretical uncertainties in the measurements of $\gamma$ and charm mixing 
at Super Flavour 
Factories and \lhcb.  Similarly measurements of decay constants and rare decays 
from data collected at charm threshold will help tune theoretical tools that will
be used elsewhere.

Other fundamental measurements that will be made include testing the CPT symmetry,
which could be violated through de-coherence of quantum correlations~\cite{Mavromatos:2008bz}, or as
the result of Lorentz violation in high energy theories~\cite{Kostelecky:1997mh}.  

In analogy with rare $B$ decays, one will be able to constrain new physics scenarios
using rare $D$ decays.  Here the advantage that \superb has over other experiments such
as \belletwo and \lhcb is the ability to cleanly extract signals, and the data sample
accumulated at \superb will be fifty times larger than that expected at BES III. 
This large data sample is particularly important when searching for rare or forbidden decays.
Finally, it is possible to make significant improvements on the precision of charm mixing
parameters by using both data from the \FourS and from the $\psi(3770)$.  

\begin{figure*}[!ht]
\resizebox{16cm}{!}{
\includegraphics[width=0.44\textwidth]{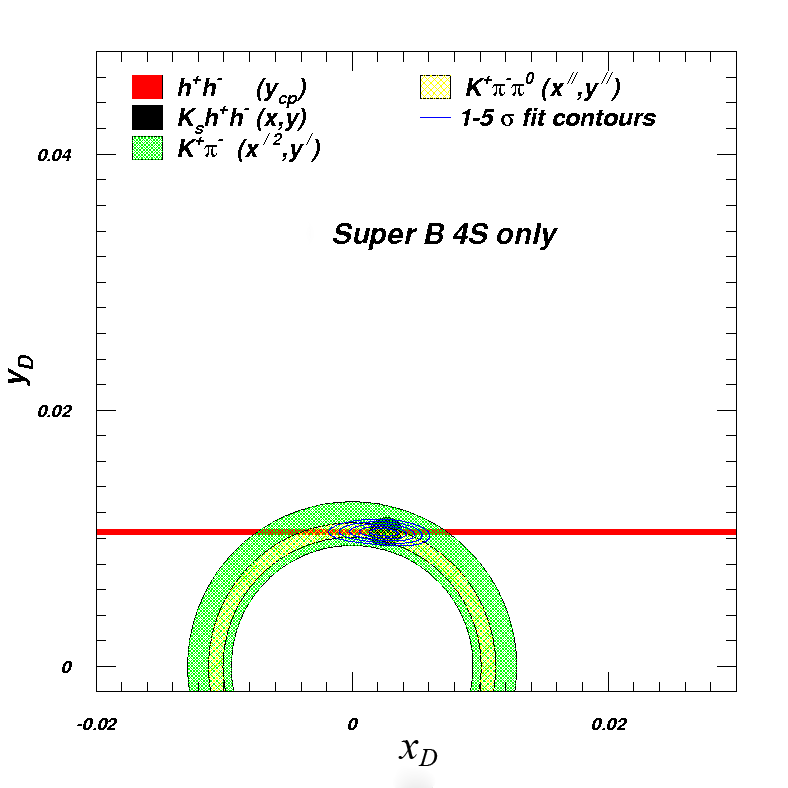}
\includegraphics[width=0.44\textwidth]{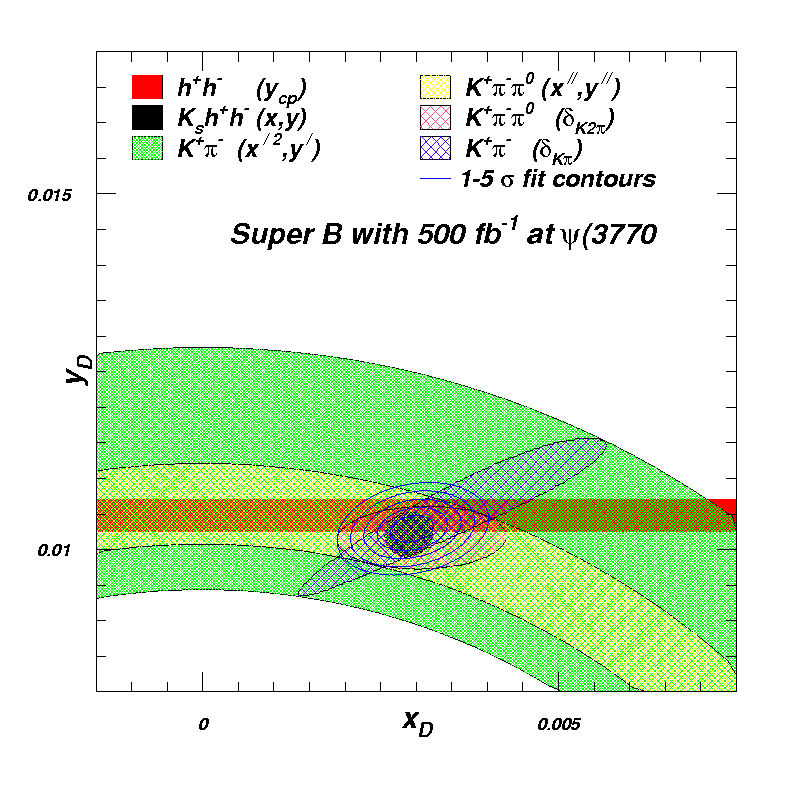}
}
        \caption{Constraints on charm mixing expected at \superb using (left) only data accumulated at the 
                 \FourS, and (right) also using data from the $\psi(3770)$.} \label{fig:bevan:charmatsuperb}
\end{figure*}

Using data from the $\psi(3770)$ and the \FourS, one would expect to be able to measure mixing parameters
in the charm sector $x_D$ and $y_D$ to a precision of $2.0\times 10^{-4}$, and $1.2\times 10^{-4}$.
The potential
of \superb for charm mixing parameters is illustrated in Figure~\ref{fig:bevan:charmatsuperb}.
More details
of this estimate can be found in Ref.~\cite{physicswp}.

\section{Precision Electroweak Physics}
\label{bevan:precisionew}

The Weinberg angle resulting from electroweak symmetry breaking has been measured
precisely at SLC and LEP~\cite{sld,lep}, through the study of $e^+e^- \to f \overline{f}$ at the $Z^0$ 
pole, where $f$ is a fermion.  Interpretation of this result as a measure of $\sin^2\theta_W$ relies on understanding 
small hadronic uncertainties at the $Z^0$.  It is possible to perform a precision measurement
of $\sin^2\theta_W$ using the left right asymmetry method employed by SLC for 
$e^+e^- \to\mu^+\mu^-$, $\tau^+\tau^-$, and $c\overline{c}$ transitions at the \FourS,
where theoretical uncertainties are negligible.  In order to do this, one must
have a polarised electron beam, as is the case for \superb.  Assuming that the electron
polarisation is 80\% and that the uncertainty on the measured polarisation is below 1\%,
then \superb should be able to perform a measurement of $\sin^2\theta_W$ with uncertainty 
of the order of $0.0002$ using di-muons.  Measurements with $\tau$ and charm quark pairs can also
be made, but with less precision. For comparison, the SLC result in Ref.~\cite{sld} is 
$\sin^2\theta_W = 0.23098 \pm 0.00026$.  Thus \superb will be able to measure this fundamental
parameter at a center of mass energy of 10.58 GeV with precision comparable to the
existing measurements at the $Z^0$.  In addition to measuring this parameter, it will be 
possible to test neutral current universality to high precision, and probe new physics
scenarios, for example models with $Z^\prime$.  More details can be found in 
Ref.~\cite{physicswp} and references therein.

\section{\boldmath $\tau$ physics programme}
\label{bevan:tau}

\superb has a broad $\tau$ physics programme ranging from searches for Lepton Flavour
Violation (LFV), and CP violation, through to a number of precision tests of the 
Standard Model, including lepton universality tests and CPT.  

Given polarised electrons, it is possible to determine the polarisation of $\tau$ leptons 
in \superb.  As a result, the reconstructed helicity angle distribution for $\tau$ decays
can be used as a discriminating variable to suppress background.  This feature of $\tau$
analyses at \superb will enable searches for the Lepton Flavour Violating decays of 
$\tau \to \ell \gamma$, where $\ell = e, \mu$ to be performed down to a level of
a few $10^{-9}$.  With regard to $\tau\to 3\ell$, the anticipated upper limits
are $2 - 8\times 10^{-10}$, depending on the three lepton final state.

With regard to other $\tau$ measurements, the polarisation enables one to measure
the $\tau$ EDM and $g-2$ parameters, with anticipated sensitivity of $17 - 34 \times 10^{-20}$ for the EDM,
and a precision of a few $\times 10^{-6}$ for $g-2$.   More details of the 
$\tau$ physics programme at \superb can be found in Ref.~\cite{physicswp}.

\section{Direct Searches and Spectroscopy}
\label{bevan:spectroscopy}

While the majority of new physics searches at \superb concentrate on the potential 
for an indirect discovery, there is a class of light new particles that may be directly
manifest in the data.  These are light scalar particles that either fall into the category of 
a light dark matter candidate, or a light Higgs particle~\cite{direct1,direct2,direct3,direct4,direct5,direct6}.  
Dark matter could be light, and have gone undetected by experiments so far, if it couples 
weakly to the $Z^0$.  Many extensions of the SM introduce 
several Higgs particles, and in extensions to MSSM it is possible for some of these
to be lighter than twice the $b$ mass.  Thus searches for both light Dark Matter (with masses less than
about 5 GeV) and light Higgs particles at \superb are essential in order to constrain their possible 
existence.

In addition to direct searches for manifestations of dark matter, it is possible to indirectly
search for possible evidence of so-called dark force, with an associated hidden sector~\cite{darkforces1,darkforces2}.  
This concept has recently emerged, and is rapidly evolving field that is being tested by 
data from both astro-particle and particle physics experiments.
In this model dark matter particles with masses less than the TeV scale can annihilate
in order to create a dark photon $A^\prime$, a gauge boson with mass of $\sim 1$ GeV. The dark photon
can then decay into SM particles, and if the mass of the $A^\prime$ is below twice the mass
of the proton, then $A^\prime$ is expected to decay into di-$e$, $\mu$, or $\pi$ final states.
As with light Higgs, and light dark matter, \superb will be instrumental in constraining 
models of dark forces.

In terms of SM spectroscopy, as was the case with the existing $B$ factories, 
\superb will be able to perform a wide range of searches for, and precision measurements
of light mesons and exotic particles. 
These studies commenced with Belle's discovery of the X(3872)~\cite{bevan:x3872}, and
a second boost to this activity was initiated by the discovery of the Y(4260) in the 
study of $J/\psi \pi\pi$ using ISR data at \babar~\cite{bevan:y4260}. 
As a result of the current plethora of 
activity there are a number of outstanding issues.  
While the masses and widths of many of these particles are now well know,
in some cases there remain issues with the determination of spin-parity 
assignments, and understanding the primary branching fractions of these.
Other outstanding issues range from simply confirming the existence of a 
particle, as is the case of the recently observed $Z^+(4430)$~\cite{belle:z4430}
to determining their underlying nature.
By understanding the underlying nature of some of these particles, one 
could make significant steps forward in our understanding of QCD, and 
in testing the framework of Lattice QCD.

\section{Elucidating new physics}
\label{bevan:archiologoy}

There are a vast number of observables that will be measured at \superb in $B$, $D$, $\tau$,
$\NS$ decays, as well as using other light mesons and baryons.  Measurements of rare
decays, that are loop, or Flavour Changing Neutral Current (FCNC) dominated can be used
to constrain many different types of new physics.  In analogy to the way that FCNCs
have shaped our understanding of the SM, these will shape understanding of new physics models. 
Precision tests of CP conjugate processes
probe phase information of the CKM mechanism in the quark sector.  In the charged lepton
sector one is able to constrain possible LFV through searches for forbidden 
transitions in $\tau$ decays.  The common coupling of charged leptons, irrespective of species,
in the SM has been tested with some precision already in a number of possible ways.  Studies
of light mesons, including $\NS$ decays to di-lepton final states can be used to increase
the precision of these constraints and further test the validity of this symmetry.  
Our physical framework is built upon the concept of Lorentz invariance, and recently
there have been attempts to go beyond this constraint in order to understand high energy
theories such as quantum gravity.  A consequence of Lorentz invariance is the conservation
of the CPT symmetry.  This symmetry can be tested in $B$, $D$, and $\tau$ decays at \superb,
to complement the tests planned and already performed at the previous generation of $B$ factories,
and in the kaon system through experiments such as CP-LEAR, KLOE, and KLOE-2.  
If a CPT violation signal were to be found, this
could have profound impact on our understanding of the foundations of physics.  Similarly any 
positive result of a direct search for dark matter candidates would have a profound impact
on the understanding of both particle and astro-particle physics.

Ultimately if no deviations from the SM are found in data collected at \superb,
then we shall be no closer to understanding the nature of new physics at high energy. 
However this outcome is not is not a bleak one, as a number of erroneous theories may have been ruled out by 
measurement, and any remaining candidate theory of new physics would be strongly 
constrained by those very measurements found to be compatible with the SM.  In turn \superb would have
performed a precision test of the electroweak and flavour sectors of the SM.

Considering the SM confirmation scenario, it is worth noting that this is 
the most versatile experiment proposed to perform precision tests of 
the CKM mechanism.  One is able to extract measurements of several of the 
matrix elements through both inclusive and exclusive analyses, measure all of the 
angles of the unitarity triangle to the level of, or better than, a degree, 
complement these measurements with information from a number of rare decays,
of both $B$ and $D$ mesons, and provide information that will help the interpretation
of some of these measurements.

Section 10 of Ref.~\cite{physicswp}, and references therein, discuss current theoretical understanding
of how one can take sub-sets of the measurements described above to discriminate between
sub-sets of possible new physics scenarios.  Phenomenological work is on going with regard
to this problem, with the ultimate goal of being able to combine measurements in a 
global way in order to satisfy particle physics and cosmological constraints of this data.

\section{Summary}

The proposed \superb experiment is a versatile Super Flavour Factory with the 
potential to provide many complementary constraints on new physics scenarios
to elucidate our understanding of physics beyond the Standard Model.  
Details of the \superb physics programme
are reported in Ref.~\cite{physicswp}. The
potential of \superb goes beyond that of the \belletwo experiment, which 
has recently had funding approved for an initial phase of accumulating data.
The ultimate goal of \belletwo is to accumulate 50\invab of data at the 
\FourS as well as investigating other \NS resonances by 2020.  The goal 
of \superb is to integrate 75\invab of data at the \FourS on a similar time scale,
and to run at the $\psi(3770)$ as well as at other \NS resonances.  If the 
nominal luminosity for \superb can be achieved using the baseline design,
it may be possible for \superb to reach a luminosity four times the design
goal subsequent to this initial phase.  

In addition to the larger integrated luminosity target of \superb, the electron
beam will be 80\% polarised.  This opens up the possibilities of being able
to measure \sinsqthetaw precisely at both the \FourS and $\psi(3770)$.  Such
a measurement at the \FourS would be theoretically clean and with a similar 
precision to the corresponding LEP result at the $Z$ pole. 

In terms of the physics programme at the $\psi(3770)$, pairs of neutral $D$ mesons will be 
created in a quantum-correlated state in analogy with $B^0$ production 
at the \FourS.  This opens up the possibility of studying time-dependent 
\CP asymmetries at charm threshold. In addition to this, one will be able to 
precisely measure decay constants and branching fractions of light $D$ mesons.
Some of these measurements will impact on theoretical uncertainties
that affect measurements within the $B$ physics programmes of the Super Flavour
Factories and \lhcb.

With regard to spectroscopy measurements, there are a number of outstanding issues
as a result of the recent work published by the current $B$ factories.  A
prime example of this is the confirmation of the $Z^+(4430)$, recently discovered by 
\belle. With one hundred times the data, these issues should be better understood.
In addition to the SM spectroscopy, \superb will also be able to place interesting constraints
on light scalar Dark Matter or Higgs scenarios, as well as Dark Forces.

The \superb experiment has the ability to measure more observables related to quark and 
lepton flavour, electroweak symmetry breaking and dark matter, than any other proposed
or existing flavour experiment.  Highlights of the physics potential reported in 
Ref.~\cite{physicswp} have been recapitulated here.  Using the measurements proposed
we will ultimately be able to strongly constrain or discover a sign of physics
beyond the SM.  In the case where one finds no deviation from the SM, then the 
resulting set of measurements obtained will place stringent constraints on 
model builders concerned with constructing a theory of particle interactions 
at high energy or in the early universe.


\begin{thebibliography}{9}

\bibitem{physicswp} B.~O'Leary {\em et al.} [\superb Collaboration], arXiv:1008.1541.

\bibitem{panta_2006} P. Raimondi in ``2nd LNF Workshop on \superb'',
Frascati, Italy, March 16-18 2006
   {\small{\tt{http://www.lnf.infn.it/conference/superb06/}}}; and in
{\it Proceedings of Particle Accelerator Conference (PAC 07)},
   Albuquerque, New Mexico, USA, June 25-29, 32 (2007).

\bibitem{ref:white_acc} ``Design Progress Report for the \superb\ Accelerator'', (2010), in preparation.

\bibitem{detectorwp} E. Grauges {\em et al.} [\superb Collaboration], arXiv:1007.4241.

\bibitem{belletworpt} Belle-II, KEK Report 04-4.
\bibitem{belletwo} T. Aushev {\em et al.}, arXiv:1002.5012.

\bibitem{chang} P. Chang, these proceedings.
\bibitem{marco} M. Ciuchini, these proceedings.

\bibitem{Altmannshofer:2009ma}
  W.~Altmannshofer, A.~J.~Buras, D.~M.~Straub and M.~Wick,
  JHEP {\bf 0904} (2009) 022
  [arXiv:0902.0160 [hep-ph]].

\bibitem{Kostelecky:1997mh}
  V.~A.~Kostelecky,
  Phys.\ Rev.\ Lett.\  {\bf 80} (1998) 1818
  [arXiv:hep-ph/9809572].


\bibitem{bevan:wolfenstein} L. Wolfenstein, Phys. Rev. Lett. {\bf 51}, 1945 (1983).
\bibitem{Mavromatos:2008bz}
  N.~E.~Mavromatos and S.~Sarkar,
  Phys.\ Rev.\  D {\bf 79} (2009) 104015
  [arXiv:0812.3952 [hep-th]].



\bibitem{sld} K. Abe et al. [SLD Collaboration], Phys. Rev. Lett. 73, 25 (1994), hep-ex/9404001.
\bibitem{lep} S. Schael et al. (LEP Electroweak Working Group), Phys. Rept. 427, 257 (2006), hep-ex/0509008.



\bibitem{direct1}
R. Dermisek and J. F. Gunion, Phys. Rev. D73, 111701 (2006), hep-ph/0510322. 
\bibitem{direct2}
R. Dermisek, J. F. Gunion, and B. McElrath, Phys. Rev. D76, 051105 (2007), hep-ph/0612031. 
\bibitem{direct3}
M.-A. Sanchis-Lozano (2007), 0709.3647. 
\bibitem{direct4}
J. F. Gunion, D. Hooper, and B. McElrath, Phys. Rev. D73, 015011 (2006), hep-ph/0509024. 
\bibitem{direct5}
B. McElrath, Phys. Rev. D72, 103508 (2005), hep- ph/0506151. 
\bibitem{direct6}
P. Fayet, Phys. Rev. D75, 115017 (2007), hep- ph/0702176. 
\bibitem{direct7}
C. Bird, R. V. Kowalewski, and M. Pospelov, Mod. Phys. Lett. A21, 457 (2006), hep-ph/0601090.

\bibitem{darkforces1} N. Arkani-Hamed, D. P. Finkbeiner, T. R. Slatyer, and N. Weiner, Phys. Rev. D79, 015014 (2009), 0810.0713. 
\bibitem{darkforces2} M. Pospelov and A. Ritz, Phys. Lett. B671, 391 (2009), 0810.1502.

\bibitem{bevan:x3872} K.~Abe {\em et al.}, Phys.\ Rev.\ Lett.\ {\bf 91} (2003) 262001.
\bibitem{bevan:y4260} B.~Aubert {\em et al.}, Phys.\ Rev.\ Lett.\ {\bf 95} (2005) 142001.
\bibitem{belle:z4430} R. Mizuk {\em et al.}. Phys.\ Rev.\ D{\bf 78}, (2008) 072004.
\end{thebibliography}
\end{document}